\documentclass[12pt,fleqn]{article}
\usepackage{psfig}
\newcommand{\beq}{\begin{equation}}
\newcommand{\eeq}{\end{equation}}
\newcommand{\beqn}{\begin{eqnarray}}
\newcommand{\eeqn}{\end{eqnarray}}
\newcommand{\bea}[1]{\beq\begin{array}{#1}}
\newcommand{\eea}{\end{array}\eeq}

\newcommand{\Z}{Z}
\newcommand{\ma}{monopole-antimonopole }
\newcommand{\dual}[1]{\left[#1\right]^d}
\newcommand{\diff}{\partial}

\newcommand{\ve}{\langle \Phi\rangle}
\newcommand{\be}{\begin{equation}}
\newcommand{\ee}{\end{equation}}
\newcommand{\preprint}{~\vspace{-1.5cm}\begin{flushright}
{\large ITEP-TH-73/98 \\
FISIST/16-98/CFIF}\end{flushright}\vspace{1.0cm}}
\begin{document}

\begin{titlepage}
\preprint

\centerline{{\bf{{SHORT STRINGS IN THE ABELIAN HIGGS MODEL}}}}

\vspace{2.0cm}
\centerline{{\bf F.V. Gubarev, M.I. Polikarpov,}}
\vspace{0.3cm}
\centerline{Institute of Theoretical and Experimental Physics,}
\centerline{
B. Cheremushkinskaya, 25, 
117259 Moscow}
\centerline{E-mail:
Fedor.Gubarev@itep.ru, polykarp@vxitep.itep.ru} \vspace{0.5cm} 
\centerline{{\bf V.I. Zakharov}}
\vspace{0.3cm}
\centerline{Max-Planck Institut f\"ur Physik,}
\centerline{F\"ohringer Ring 6, 80805 M\"unchen, Germany.}
\centerline{ E-mail: xxz@mppmu.mpg.de} 
\vspace{1.5cm}

\begin{abstract}
We consider the monopole-antimonopole static potential in the
confining phase of the Abelian
Higgs model and in particular the corrections to the Coulomb-like
potential at small distances $r$. 
By minimizing numerically the classical energy functional
we observe a linear in $r$, stringy correction 
even at distances much smaller than the apparent physical scales.
We argue that this term is a manifestation of
the condition that the monopoles are connected by
a mathematically thin line along which the scalar field
vanishes. These short strings modify the operator product
expansion as well. Implications for QCD are
discussed.
\end{abstract}
\end{titlepage}

 ``String'' or vortex-like configurations play an important role in a
number of areas of physics and in theoretical speculations. The  best
known perhaps is the flux tube of superconductivity (``Abrikosov string'')
and its relativistic version in particle physics, in the Abelian Higgs
model(``Nielson-Olesen string'') \cite{ano}. 
Such configurations are also believed to
play an important role in  QCD, and the  confining force between two
quarks is often described as due to the stretching of the string
\cite{nambu}.

Such physical strings have a definite thickness, reflecting the
balance of the various forces going into their composition, and this
introduces a certain length scale into the problem. For example for
the forces between two static sources, like the quarks of QCD or
between hypothetical magnetic monopoles in a superconductor, the
string gives a linear potential energy at large distances, reflecting
a certain energy per unit length of the string.

Naturally at small distances, that is distances small compared to the
thickness of the string, where the picture of an energy per unit
length does not seem to apply, we would expect the term linear in the
separation to disappear. In this paper we would like to point that,
surprisingly, even at small distances, that is less than the thickness
of the string, there remains a stringy term in the energy.  This term
may be interpreted as physical manifestation of the mathematical Dirac
string \cite{dirac} which accompanies such objects as magnetic
monopoles.

In the bulk of the paper we will consider the Abelian Higgs model 
(AHM) while implications for QCD are summarized in the concluding remarks.
The AHM describes  
a gauge field $A_{\mu}$ interacting with a charged scalar field $\Phi$
as well as self-interactions of the scalar field,
and
the corresponding action is:
\be\label{AHM_action}
S= \int d^4x \left\{
\frac{1}{4e^2} F^2_{\mu\nu} + \frac{1}{2} |(\diff - i A)\Phi|^2 + 
\frac{1}{4} \lambda (|\Phi|^2-\eta^2)^2
\right\}
\ee
where $e$ is the electric charge, $\lambda,\eta$ are constants and
$F_{\mu\nu}$ is the electromagnetic field-strength tensor,
$F_{\mu\nu}\equiv\diff_{\mu}A_{\nu}-\diff_{\nu}A_{\mu}$.
The scalar field 
condenses in the vacuum,
$\ve =\eta$ and the physical 
vector and scalar particles are massive, $m^2_V=e^2\eta^2, m_H^2= 2
\lambda \eta^2$.

The model is famous 
to provide with a relativistic analog of
the Landau-Ginzburg theory of superconductivity. In particular,
if one introduces a \ma pair as an external probe its static potential
$V(r)$ grows linearly with the distance $r$ at large $r$:
\be
lim_{r\rightarrow \infty}V(r)~=~ \sigma_{\infty} r\label{growth}
\ee
The growth of the potential (2) is well understood in terms of 
the Abrikosov-Nielsen-Olesen (ANO) strings \cite{ano}.
These strings are solutions to the classical equations of motion
corresponding to the action (1) and carry (quantized) magnetic flux
equal to
$2\pi n/e$.
Because of this, strings may end up with \ma pairs.
The salient features of the solution is that the Higgs 
field $\Phi$ disappears on the axis of the string
and reaches its vacuum value at transverse distances of order $1/m_H$
while the magnetic field extends to distances of order $1/m_V$.
In the limit of large distances, $r\gg m_V^{-1},m_H^{-1}$
the ANO string can be considered as thin
and the constant $\sigma_{\infty}$ in Eq. (\ref{growth}) is then its 
tension.

We will study the static energy of a \ma pair at short distances,
$r\ll m_V^{-1}, m_H^{-1}$. 
The potential $V(r)$ is then Coulomb like because of
exchange of the photon.
As is mentioned above, the ANO strings are not relevant at such distances,
and one would expect that the power corrections to the Coulomb potential
are simply due to a non-vanishing vector particle mass. That is, up to a 
constant which can be included  into definition of the heavy masses:
$$\lim_{r\to 0}{V(r)_{Yu}}~ \approx~ 
- \frac{\alpha_{mag}}{r}\left(1 +\frac{m_V^2 r^2}{2}\right).$$
Note the change of the sign in front of the linear term 
as compared
to the stringy potential (\ref{growth}).

The central point of the present letter is  
that these naive expectations are not true because,
even at scales when the ANO string is irrelevant, there still exists an
infinitely thin line inherent to the problem. 
Namely, there is a
line connecting
the monopoles along which $\Phi=0$.
Its existence follows from the  observation \cite{thooft}
that the world sheets with $\Phi=0$ are either closed or have 
monopole trajectories as their boundaries. In other words, 
it is a topological condition that magnetic charges are connected by
a line $\Phi=0$, no matter how small the distance $r$ between the charges is.

Because this topological condition is so important for further results,
it is worth noting that it can be rederived in the
language of the Dirac string \cite{dirac} as well.  
Namely, the infinitely thin line discussed above is nothing
else but the Dirac string connecting the monopoles. The possibility of
its dynamical manifestations arises from the fact that the Dirac
string cannot coexist with $\Phi\neq 0$ and $\Phi$ vanishes along the
string.  Indeed, the self-energy of the Dirac string, is normalized to
be zero in the perturbative vacuum. To justify this one can invoke
duality and ask for equality of self-energies of electric and magnetic
charges. Since the electric charge has no string attached the
requirement would imply vanishing energy for the Dirac string.
However, if the Dirac string would be 
embedded into a vacuum with
$\ve\neq 0$ then its energy would jump to infinity since there is the
term $1/2|\Phi|^2A_{\mu}^2$ in the action and $A_{\mu}^2\rightarrow
\infty$ for a Dirac string.  Hence, $\Phi=0$ along the string and it
is just the condition found in \cite{thooft}.  In other words, Dirac
strings always rest on the perturbative vacuum which is defined as a
vacuum state obeying the duality principle. Therefore, even in the
limit $r\to 0$ there is a deep well in the profile of the Higgs field
$\Phi$. This might cost energy which is linear with $r$ even at small
$r$.

By solving numerically the classical equations of motion with a
boundary condition $\Phi=0$ along the line connecting the monopoles we
do find a linear stringy piece,
i.e. with a positive coefficient in front of $r$, 
in the potential even in the limit
$r\to 0$. Another manifestation of the fact that we are
dealing with a kind of elementary string is the breaking of the operator
product expansion.  Indeed the standard operator product expansion
(OPE) assumes that short distance expansions can be derived in terms
of exchange of elementary particles.  On the other hand, the topolgical
string $\Phi=0$
cannot be constructed in terms of particle exchanges but should
be added independently. Since the effective string tension for small
size strings is proportional to $<\Phi>^2$, the standard OPE breaks
down at the level of $r^2<\Phi>^2$ corrections to pure
perturbative results.

To evaluate the static energy $E(r)$, we first need to introduce the
monopole trajectories explicitly into the path integral: 
\beq
E(r)=\lim\limits_{T\to\infty} 
\left(-\frac{\diff}{\diff T}\right)\ln H(C) \, , 
\eeq 
\beq\label{t'Hooft-loop-average}
H(C)=\frac{\displaystyle 1}{\displaystyle \Z} \int DA D\Phi [D\Phi]
H(A,\Sigma^C) e^{-S(A,\Phi)} \, , 
\eeq 
here the 't Hooft loop is
defined as follows: 
\beq\label{t'Hooft-loop} 
H(A,\Sigma^C)=\exp
\left\{ \frac{1}{4e^2} \int d^4x \left[ F^2_{\mu\nu} - (F_{\mu\nu} +
2\pi\dual{\Sigma^C}_{\mu\nu})^2 \right]\right\} 
\eeq 
where $\Sigma^C$
is an arbitrary surface having the contour $C$ as a boundary,
$\delta\Sigma^C = C$.  A particular form of the contour $C$ is
determined by the monopoles trajectory and for a static \ma pair
located at distance r the contour $C$ is a rectangular loop $T \times
r$ with $T \gg r$.  In~Eq (\ref{t'Hooft-loop}) $\dual{...}$ denotes
the duality operation so that for any tensor $T_{\mu\nu}$ of the
second rank
$\dual{T_{\mu\nu}}=\frac{1}{2}\varepsilon_{\mu\nu\lambda\rho} \;
T_{\lambda\rho}$. It can be shown that the expectation value of the 't
Hooft loop (\ref{t'Hooft-loop-average}) does not depend on the
particular choice of the surface $\Sigma^C$. 
Moreover it can be shown
\cite{gubarev} that in the string representation of the Abelian Higgs
model we have to sum over all surfaces $\sigma_C$ spanned on the loop
$C$ and also over all closed surfaces $\sigma$ (representing
virtual glueballs). The classical solution corresponds to the
minimal area surface $\sigma_C$.
Here we consider the classical approximation, hence the problem of
finding $E(r)$ is equivalent to solving classical equations of motion
with boundary condition $\Phi=0$ along the straight line connecting
the monopoles.  

Let us note that the numerical results for the
potential $V(r)$ can be found in a number of papers
\cite{numerical,suzuki90} and in this respect our approach is not new.
However, there are no measurements
dedicated specifically to small corrections to the Coulombic potential
at $r\rightarrow 0$. Indeed the standard question addressed
is how the Coulomb-like behaviour of the potential at short
distances is transfromed into the linear potential at large
distances (see, e.g., Ref. \cite{baker} and references therein).
It is only very recently that it was recongnized that the power
correction to the potential at short distances in QCD
could signify a new phyisics \cite{az}. Hence, there are no measurements
which would provide error bars of the slope of the potential at small distances.
Moreover, one could argue that at small $r$ the strong magnetic  
field ``pushes out'' the Higgs field in any case and, therefore, the 
potential energy is insensitive to the condition $\Phi=0$ along
a line connecting the magnetic charges. 
Since in answering this kind of questions
we rely on the numerical results, we need dedicated measurements.
   
We will consider the unitary gauge, $Im\Phi=0$.
Then
the most general ansatz 
for the fields consistent with the symmetries of the problem is:
\bea{ll}
\Phi=\eta f(\rho,z)  &
A_{a}=\varepsilon_{a b} \hat{x}_b A(\rho,z)~~~~~
A_{0}=A_{3}=0 \\
\\
\rho=[x_a x_a ]^{1/2} & z=x_3 \qquad \hat{x}_a=x_a/\rho 
\qquad a=1,2 \ .
\eea
Let us introduce also a new variable
$\kappa=\sqrt{2\lambda}/e =  m_H/ m_V$ and measure all dimensional
quantities in terms of $m_V = e\eta$.  Then the energy functional is:
\bea{l}\label{energy-1}
E(r)=\frac{\pi}{e^2} \int\limits^{+\infty}_{-\infty}dz 
\int\limits^{+\infty}_{0} \rho d\rho
\left\{
[\frac{1}{\rho}\diff_\rho (\rho A)+\Sigma]^2 + [\diff_zA]^2+
\right.  
\\
\left.\rule{0.3\textwidth}{0.0mm}
+[\diff_\rho f]^2 + [\diff_z f]^2 + f^2A^2 + \frac{1}{4}\kappa^2(f^2-1)^2
\right\}
\eea
\beq\label{Sigma-1}
\Sigma=\frac{1}{\rho}\delta(\rho) \cdot \int\limits^{1}_{-1} d\xi \; \delta(z-\xi\frac{r}{2})
\eeq

In the limit $r \to 0$ the Coulombic contribution becomes singular. 
The easiest way to separate the singular piece is to change the
variables $A=A_d+a$, where $A_d$ is the solution
in the absence of the Higgs field:
\be\label{dipole_potential}
A_d=\frac{1}{2\rho}\left[
\frac{z_{-}}{r_{-}} - \frac{z_{+}}{r_{+}}
\right],~~~
z_{\pm}=z \pm r/2 \qquad r_{\pm}=\left[ \rho^2 + z_{\pm}^2 \right]^{1/2}
\ee
Upon this change of variables the energy functional takes the form:
\bea{l}\label{energy-2}
E(r)=E_{self}-\pi/r + \tilde{E}(r),
\\
\\
\tilde{E}(r)=\frac{\pi}{e^2}
\int\limits^{+\infty}_{-\infty}dz \int\limits^{+\infty}_{0} \rho d\rho
\left\{
[\frac{1}{\rho}\diff_\rho (\rho a)]^2 + [\diff_za]^2 +
\right.
\\
\left.\rule{0.3\textwidth}{0.0mm}
+[\diff_\rho f]^2 + [\diff_z f]^2 + f^2(a+A_d)^2 + \frac{1}{4}\kappa^2(f^2-1)^2
\right\}
\eea

We investigated numerically $\tilde{E}(r)$ in the region $0.1 < r < 5.0$
(details of the procedure will be given elsewhere).
Fig.~1 represents the results of our computations for various
$\kappa$ values and clearly demonstrates that there is a linear
piece in the potential even in the limit $r\ll 1$.
While obtaining the values of the slope 
$\sigma_0$ at small distances we separated the Yukawa-potential
contribution by fitting the energy as follows:
\beq\label{fitting-function}
\tilde{E}_{fit}(r)= C_0 \left(\frac{\displaystyle 1-e^{-r}}{r}-1\right) + 
(\sigma_0 + \frac{1}{2}C_0)r =
\sigma_0 r + O(r^2)
\eeq
The slope $\sigma_0$ depends smoothly on the value of $\kappa$,
see Fig. 2. The linear piece in the potential at small distances 
reflects the boundary condition that $\Phi=0$
along the straight line connecting the
monopoles. To illustrate this point we show in Fig.~3
the function $(1-f(\rho,z))$ which makes the impact of the stringy
boundary condition visible.
It is noteworthy that in the Bogomolny limit ($\kappa=1$)
the slope of the linear potential within the error bars is the same 
for $r\to\infty$ and $r\to 0$.

As it is mentioned above, the existence of short strings is manifested
also through breaking of the standard operator expansion. Indeed, above we
found the potential in the classical approximation. In this
approximation the potential is usually directly related to the
propagator $D_{\mu\nu}(q^2)$ in the momentum space,
\be V(r)=\int d^3r \; e^{i{\bf q\cdot r}}\;D_{00}({\bf q}^2)
\label{aa}\ee
Moreover, as far as $q^2$ is in
Euclidean region and much larger than the mass parameters,
the propagator $D_{\mu\nu}(q^2)$ 
can be evaluated by using the OPE.
Restriction to the classical approximation implies that loop contributions
are not included. However, vacuum fields which are soft on the
scale of ${\bf q}^2$ can be consistently accounted for in this way
(for a review of see, e.g., \cite{novikov}).
This standard logic can be illustrated by an
example of the photon propagator connecting two electric currents.
Modulus longitudinal terms, we have: 
\be \label{prop}
D_{\mu\nu}(q^2) ~=~\delta_{\mu\nu}
\left({1\over q^2}+{1\over q^2}e^2\langle \Phi^2\rangle {
1\over q^2}+{1\over q^2}e^2\langle \Phi^2\rangle {
1\over q^2}e^2\langle \Phi^2\rangle {
1\over q^2}+...\right)~=~{\delta_{\mu\nu}\over q^2-m_V^2}.       
\ee
Thus, one uses first the general OPE assuming $|q^2|\gg e^2\Phi^2$
then substitutes the vacuum expectation of the Higgs field 
$\Phi$ and upon summation of the whole series of the power corrections
reproduces
the propagator of a massive particle. 
The latter can also be obtained 
by solving directly the classical equations of motion.

This approach fails, however, 
if there are both magnetic and electric charges 
present. In this case, one can choose the Zwanziger formalism
\cite{zwanziger} and work out an expression for propagation of a photon 
coupled to magnetic currents following 
literally the same steps as in (\ref{prop}), (see e.g. \cite{balachandran}).
In the gauge $n_{\mu}D_{\mu\nu}=0$ the result is::
\be \label{wrong}
\tilde{D}_{\mu\nu}(q,n)~=~{1\over q^2-m_V^2}\left(\delta_{\mu\nu}-
{1\over (qn)}(q_{\mu}n_{\nu}+q_{\nu}n_{\mu})+{q_{\mu}q_{\nu}\over(qn)^2}+
{m_V^2\over (qn)^2}(\delta_{\mu\nu}n^2-n_{\mu}n_{\nu})\right)   
.\ee
Here the vector $n_{\mu}$ is directed along
the Dirac strings attached to the magnetic charges 
and there are general arguments
that there should be no dependence of physical
effects on $n_{\mu}$ \cite{zwanziger}. On the other hand, 
if the potential energy is given by the Fourier transform of (\ref{wrong})
then its dependence on $n_{\mu}$ is explicit,
see, e.g., \cite{suzuki,suganuma} and we addressed this problem in Ref. 
\cite{gubarev}.

Note that Eq. (\ref{wrong}) immediately implies that the standard OPE
does not work any longer on the level of $q^{-2}$ corrections.
Indeed, choosing $q^2$ large and negative does not guarantee now that
the $m_V^2$ correction is small since the factor $(qn)^2$ in the
denominator may become zero. Of course, appearance of the poles in 
$(qn)$ variables in longitudinal pieces is not dangerous since they
drop due to the current conservation. However, the term proportional
to $m_V^2$ in (\ref{wrong}) cannot be disregarded and contribute, 
in particular, to (\ref{aa}).

The reason for the breaking of the
standard OPE is that even at short distances the dynamics of the short
strings should be accounted for explicitly. In particular, in the
classical approximation the string lies along the straight line
connecting the magnetic charges and affects the solution through the
corresponding boundary condition, see above. More generally, the OPE
allows to account for effect of vacuum fields, in our case for $\ve
\neq 0$. The OPE is valid therefore as far as the probe particles do
not change the vacuum fields drastically and the unperturbed vacuum
fields are a reasonable zero-order approximation.  In our case,
however, the Higgs field is brought down to zero along the string and
this is a nonperturbative effect. Thus, the stringy piece in the
potential $V(r)$ at $r\to 0$ is a nonperturbative correction which is
associated with short distances and emerges already on the classical
level.

A few remarks on the implications of the results obtained to QCD are
now in order.  To begin with, there exist detailed numerical
simulations on the lattice which confirm the dual-superconductor
picture of the confinement (for review and references see
\cite{polikarpov}).  In particular, the potential $V(r)$ at large $r$
is well described by the model \cite{baker}.  Moreover, if the
simulations are performed in the $U(1) $ projection of QCD,
condensation of a scalar field $\Phi_{mag}$ with magnetic charge
is confirmed and, moreover, the structure of the observed string which
determines the $\bar{Q}Q$ potential at large distances is well
described by the {\it classical} Landau-Ginzburg equations
\cite{BSS98}. What is especially important for us, is that the definitions
of the magnetic charges in the Abelian projection
of gluodynamics \cite{thooft} are in fact local and, therefore, the results obtianed within
the AHM with a local field $\Phi$ can imitate gluodynamics. 

Note that, naively, existence of a $U(1)$ gauge invariant
operator $|\Phi_{mag}|^2$ of dimension $d=2$ in the abelian projection
of QCD would imply infrared sensitive corrections of order $\langle
\Phi_{mag}\rangle^2/q^2$ which are calculable via the OPE.  On the
other hand, it is well known (for references see, e.g.,
\cite{novikov}) that such corrections are not present in QCD.  The
paradox is resolved through the observation that the $U(1)$ projection
of QCD is similar to the Higgs model and, therefore, the OPE breaks
down in this projection.

Moreover, we have learned that in the Higgs model there emerges in
fact a non-perturbative short distance correction of order $q^{-2}$
manifested through the slope
$\sigma_0\neq 0$.  Although the Dirac strings are specific for the abelian
projection, the results for the slope $\sigma_0$ which is a physical
quantity should be true for any gauge fixing.  In
other projections, the linear potential at small distances arises if
there are short-distance correlations in non-perturbative fields
\cite{az}.
There are attempts to extract phenomenological consequences
from hypothetical existence of short strings in QCD \cite{cnz}.
On the other hand, the most common picture of non-correlated finite
size non-perturbative fluctuations results in $\sigma_0=0$
\cite{balitsky}.

It is amusing therefore that the lattice simulation \cite{bali2} do
not show any change in the slope of the $\bar{Q}Q$ potential as the
distances change from the largest to the smallest ones available. In
the notations introduced above, 
\be \label{equal}
\sigma_{\infty}~\approx~\sigma_0.  
\ee 
Moreover, it is known from phenomenological analisis and from the
calculations on the lattice \cite{suzuki,Blimit,BSS98} that the
realistic QCD corresponds to the case $\kappa\approx 1$ where $\kappa
=m_H/m_V$. It is remarkable that, as is mentioned above, the AHM in
the classical approximation also results in the relation (\ref{equal})
for $\kappa\approx 1$.  Thus we see that existing data \cite{bali2} on
the behavior of the $\bar{Q}Q$ potential at {\it small} distances
agree with the classical approximation to AHM. Also, our results 
support indirectly phenomenological attempts to account for the novel
$1/q^2$ corrections \cite{cnz}

To summarize, we have demonstrated that the potential of a \ma pair at
distances much smaller than the inverse masses $m^{-1}_{V,H}$ does
contain a linear piece $\sigma_0 r$ with a positive $\sigma_0$.  This
linear piece is a dynamical manifestation of the topological condition
that the scalar field $\Phi$ vanishes along a line connecting the
magnetic charges. These short strings are responsible also for
breaking of the standard OPE on the level of $1/q^2$ corrections.
Note that usually the Abelian Higgs model plays the role of the
effective infrared model of gluodynamics (see reviews
\cite{polikarpov}). It is amusing that the behaviour of the $Q \bar{Q}$
potential at {\it small} distances obtained via the lattice
simulations agree with the dual-superconductor model with
$\kappa\approx 1$.

We are thankful to M.N.~Chernodub, V.A.~Rubakov, L.~Stodolsky and
T.~Suzuki for interesting discussions, F.V.G.  and M.I.P.  feel much
obliged for the kind hospitality extended to them by the staff of
Max-Planck Institut fuer Physik (Munich), and by the staff of Centro
de F\'\i sica das Interac\c c\~oes Fundamentais, Edif\'\i cio
Ci\^encia, Instituto Superior T\'ecnico (Lisboa). This work was
partially supported by the grants INTAS-RFBR-95-0681, INTAS-96-370,
RFBR-96-15-96740 and RFBR-96-02-17230a.

\newpage
\begin{figure}[ht]
\centerline{\psfig{file=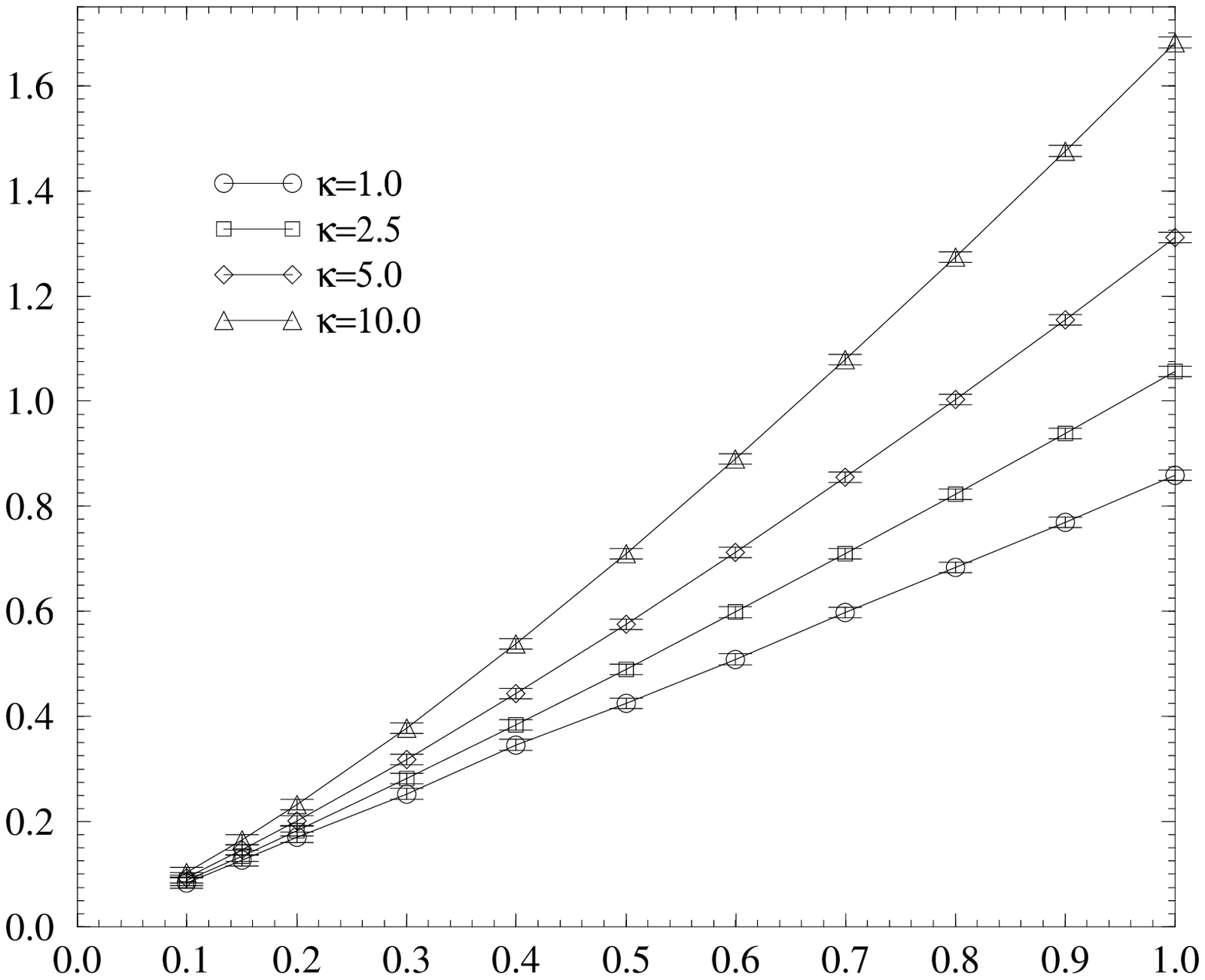,width=0.8\textwidth,silent=}}
\caption{
$\frac{\displaystyle e^2}{\displaystyle \pi}\cdot\frac{\displaystyle E}
{\displaystyle m_v}$ as a function of $m_v r$.}
\end{figure}
\begin{figure}[bht]
\centerline{\psfig{file=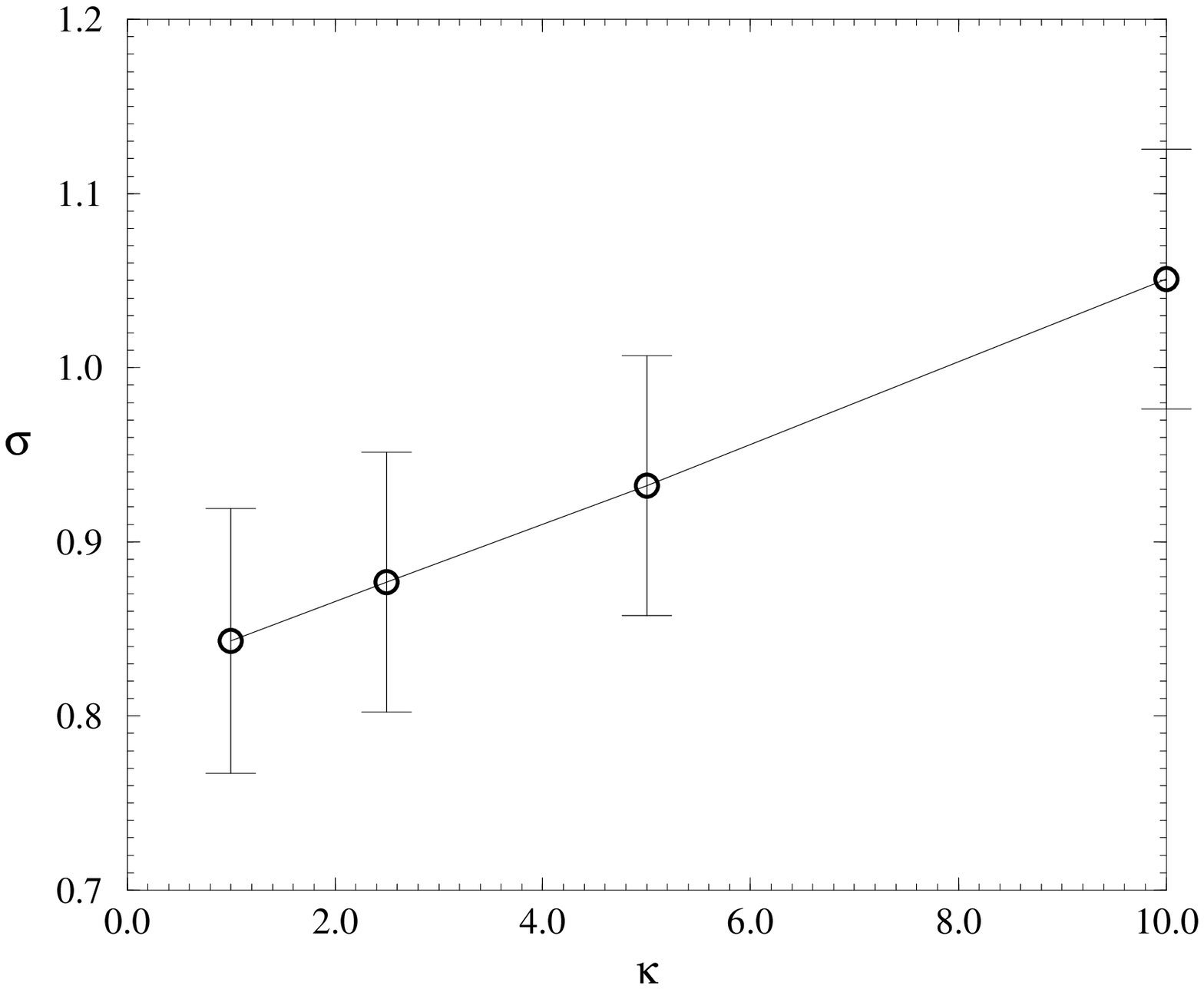,height=0.4\textwidth,silent=}}
\caption{
The linear slop $\sigma$ of the energy
$\frac{\displaystyle e^2}{\displaystyle \pi}\cdot\frac{\displaystyle E}
{\displaystyle m_v}$ in the limit $r\to 0$
as a function of $\kappa$ (see~(\ref{fitting-function})).}
\end{figure}
\begin{figure}[ht] 
\centerline{\psfig{file=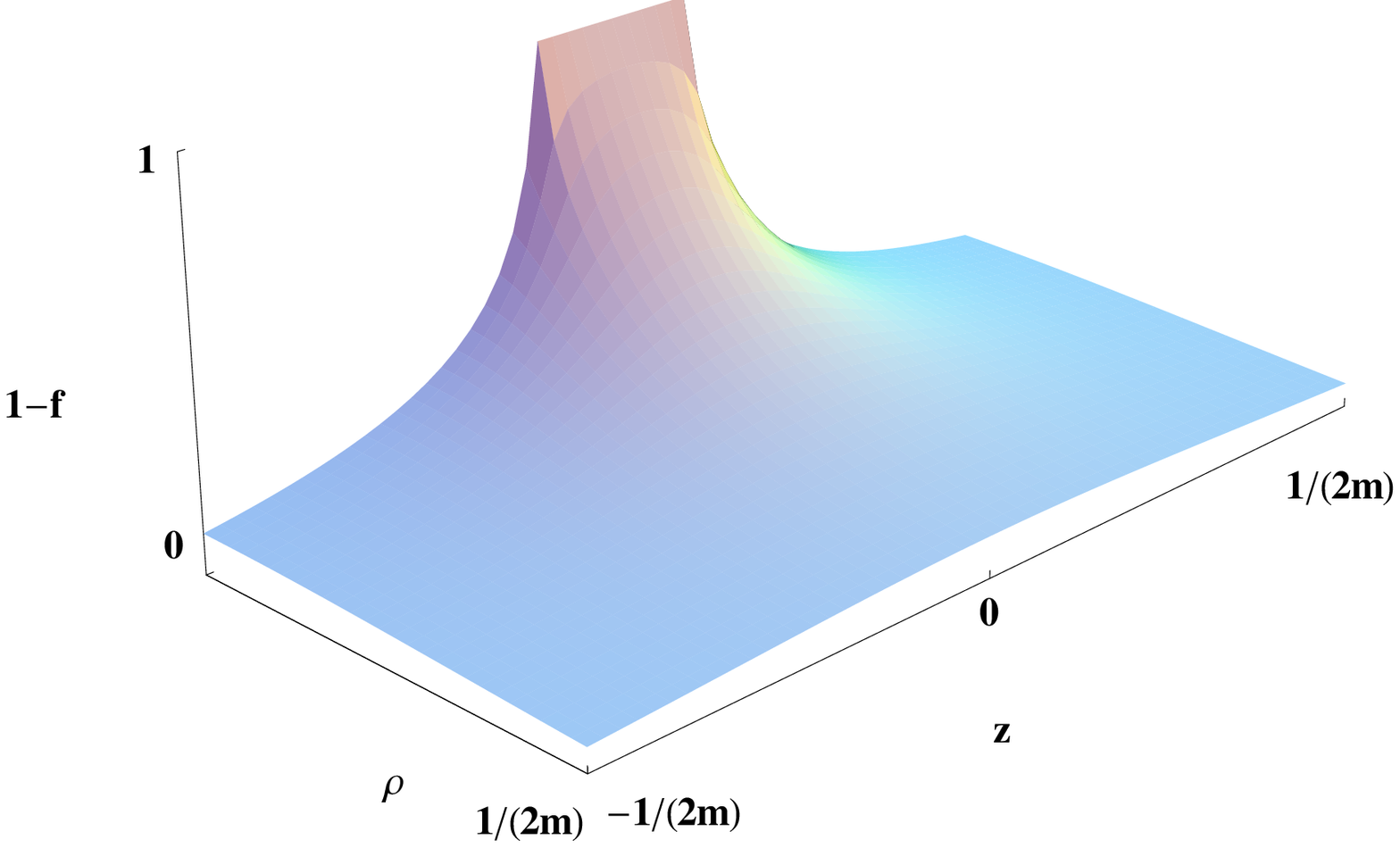,width=0.8\textwidth,silent=,clip=}}
\caption{
The function $(1-f(\rho,z))$ in the $\rho - z$ plane 
for $m_H = m_V = m$, $r = 0.2/m$. The line at which $f = 0$
is clearly seen.
}
\end{figure}

\end{document}